\makeatletter\let\ifGm@compatii\relax\makeatother
\documentclass[pre,aps,twocolumn,floatfix,superscriptaddress]{revtex4-1}
\usepackage{eso-pic,calc}
\usepackage{graphicx}
\usepackage{amsmath, amsthm, amssymb}
\usepackage{epsfig}
\usepackage{latexsym}
\usepackage{bm}

\usepackage[colorlinks=True]{hyperref}  
\hypersetup{
	colorlinks=true,  
	linkcolor=blue,  
	citecolor=blue, 
	filecolor=magenta,   
	urlcolor=blue         
}

\def\beqr{\begin{eqnarray}}
	\def\eqnr{\end{eqnarray}}
\def\beq{\begin{equation}}
	\def\bc{\begin{center}}
		\def\ec{\end{center}}
	\def\eqn{\end{equation}}

\def\ie{i.e.,~}

\begin{document}
\title{Scaling features in the Olami-Feder-Christensen model }

\author{Naveen Kumar}
\affiliation{Govt. College for Women, Prade Ground Jammu, Jammu and Kashmir 180001, India}

\author{Rahul Chhimpa}
\affiliation{Department of Physics, Institute of Science,  Banaras Hindu University, Varanasi 221 005, India}

\author{Avinash Chand Yadav}\thanks{jnu.avinash@gmail.com}
\affiliation{Department of Physics, Institute of Science,  Banaras Hindu University, Varanasi 221 005, India}

\begin{abstract}
{We consider the Olami-Feder-Christensen (OFC) model on a square lattice with open boundary conditions. The model exhibits self-organized criticality and explains the Gutenberg–Richter law observed for earthquakes. A parameter \(\alpha\) controls the level of local dissipation: \(\alpha<0.25\) corresponds to locally dissipative and \(\alpha = 0.25\) marks locally conservative dynamics. The avalanche size distribution follows a decaying power-law, with a non-universal critical exponent. Here, we examine the local and total stress fluctuations in the OFC model for both locally conservative and dissipative dynamics. The finite-size scaling analysis of the power spectra for the stress fluctuations reveals qualitatively the same but quantitatively significantly different behavior. The dynamic exponent describing the divergence of the correlation time with system size changes from nearly ballistic in the conservative to diffusive in the locally dissipative dynamics with \(\alpha = 0.21\). The local stress also exhibits a signature of nearly canonical \(1/f\) noise in the intermediate regime, and \(1/f^2\)-type scaling dominates the high-frequency regime. We further examine the probability distribution of the difference between avalanche size and area. We find a power-law behavior with a scaling exponent close to one in the conservative OFC model. The scaling feature vanishes even for the physically relevant case \(\alpha = 0.21\). To examine the robustness of such features, we also examine the same quantity in the Bak-Tang-Wiesenfeld and Manna sandpile models on a square lattice. We find that the power-law behavior survives for these systems due to locally conservative dynamics. }
\end{abstract}
\maketitle

\section{Introduction}
The Olami-Feder-Christensen (OFC) model~\cite{ofc_1992} of seismic activities remains one of the extensively studied prototypical systems of self-organized criticality (SOC)~\cite{btw_1987, Bak_1996, Christensen_2005, dhar_2006, Pruessner_2012, markovic_2014, Watkins_2016}. The SOC implies the spontaneous organization of slowly driven nonequilibrium systems that dissipate instantaneously into a critical state characterized by critical avalanches and long-range temporal scaling features. Diverse systems exhibit SOC, for example, sandpiles~\cite{btw_1987} to earthquakes~\cite{ofc_1992, bak_2002, ida_2024} and biological evolution~\cite{bs_1993, singh_2023, bshd_2024} to neuronal circuits~\cite{Beggs_2003, plenz_2014, wang_2016}.

Typically, the critical avalanche characteristics such as size \(S\), area or spatial extent \(A\), and duration \(T\) observed in SOC systems follow a power-law decaying probability distribution \(P(X, L)\sim X^{-\tau_X}\) with an upper cutoff that diverges with the system size \(L^{D_X}\). If the finite-size scaling (FSS) applies, we can write a scaling ansatz~\cite{Jensen_1998, Pruessner_2012}
\begin{equation}
	P(X, L) \sim L^{\tau_X D_X}F(X/L^{D_X}),
	\label{eq_pd_1}
\end{equation}
where the scaling function in Eq.~(\ref{eq_pd_1}) varies as \(F(v) \sim v^{-\tau_X}\) for \(v \ll 1\). Also, a scaling relation \(\gamma_{ST} = (\tau_T-1)/(\tau_S-1)\) describes scaling behavior of average size with duration.

For the dissipative OFC model on a square lattice, numerical results previously showed that the avalanche size distribution follows a power-law behavior with a non-universal exponent~\cite{ofc_1992, grassberger_1994}. Specifically, \(\tau_S \approx 1.8\) corresponds to a physically relevant case arising at the dissipation parameter value \(\alpha = 0.21\). However, the distribution violates ordinary FSS. Lise and Paczuski showed that subsystems of linear dimension obey finite (subsystem) size scaling with universal exponents \(D_S = 2\) and \(\tau_S = 1.8\), independent of the dissipation parameter \(\alpha\)~\cite{lise_2001}. Lise and Paczuski studied the dissipative OFC model on a random network. They found a signature of criticality, obeying FSS with universal critical exponents \(D_S = 1\) and \(\tau_S \approx 1.65\), independent of the parameter \(\alpha\)~\cite{lise_2002}. Regarding random graphs as a high-dimensional limit of a regular lattice, the OFC model on it represents mean-field behavior. The dissipative OFC model on a small world topology with \(\alpha = 0.21\) and \(p = 0.00586\) (rewiring probability) reveals a clear data collapse with \(D_S = 2\) and \(\tau_S = 1.8\)~\cite{tadic_2006}. The difference between two successive avalanches with a constant time shift resembles an intermittent process with a non-Gaussian probability at criticality, supported by empirical data. Davidsen and Schuster studied the power spectra of the avalanche signals in the two-dimensional (2D) OFC model, revealing that the signals can show a signature of \(1/f\) noise when the threshold stress itself evolves as a random walk~\cite{davidsen_2000}.

This paper reports two scaling features for the OFC model on a square lattice with open boundary conditions. First, to understand the underlying temporal correlation in the OFC model, we examine the power spectra for the local and total stress fluctuations as a function of the drive time scale for both locally conservative and dissipative dynamics. While the stress fluctuations exhibit \(\sim 1/f^2\)-type scaling in the high-frequency regime, we find a signature of nontrivial \(1/f\) noise in the intermediate frequency regime for the local stress signals. Interestingly, FSS reveals interesting scaling properties. The correlation time diverges with system size, with a dynamic exponent \(z\) that changes from nearly ballistic to diffusive behavior, from locally conservative to dissipative with \(\alpha = 0.21\), respectively.

Secondly, while the avalanche size and area follow a power-law probability even for the locally dissipative OFC model, we show that the probability distribution of the difference between avalanche size and area \(X = S-A\) exhibits a power-law decaying behavior with a critical exponent nearly equal to one for the locally conservative dynamics. The scaling behavior vanishes when the local dissipation level increases. We also check the robustness of the scaling features in the Bak-Tang-Wiesenfeld (BTW) and Manna sandpile models on a square lattice. In both systems, the scaling behavior persists because of locally conservative dynamics.

As suggested in previous studies, the earthquake seismic moment is related to the area nonlinearly~\cite{landes_2016}. A recent study of large earthquakes shows that complex rupture dynamics are common, and fault heterogeneity leads to multiple rupture episodes~\cite{wong_2026}. In the OFC model, the size of the avalanche corresponds to the earthquake's magnitude, and the area (the number of distinct toppling sites) can be related to the spatial extent of the rupturing fault. The difference \(X=S-A\) seems an interesting quantity as it can isolate the excess activity during an earthquake. It can help us classify avalanches by their internal complexity rather than their overall size.

The plan of the paper is as follows. In Sec.~\ref{sec_2}, we begin by presenting the OFC model and quantities of interest. In Sec.~\ref{sec_3}, we study the power spectral density of local and total stress fluctuations in conservative and dissipative OFC dynamics by applying the FSS method. In Sec.~\ref{sec_4}, we examine the probability distribution of the difference between avalanche size and area with different system sizes and the effect of the dissipation parameter. Finally, we conclude the paper with an outlook in Sec.~\ref{sec_5}.

\section{OFC model}{\label{sec_2}}
Earthquakes can arise from stick-slip dynamics in the Earth’s crust sliding along faults. In this context, the OFC model of earthquakes corresponds to a simplified cellular automata version of the Burridge-Knopoff system~\cite{BK_1967, carlson_1989, carlson_1991}.
The algorithm used to simulate the OFC model is as follows. We consider a square lattice with linear extent \(L\), with open boundary conditions. Let \(F_i\) be a local force or stress variable, where \(i\) runs from 1 to \(L^2\). Initially, we assign a random value to each \(F_i\) drawn from a uniform probability density function in an interval \([0, F_0]\). In simulations, we set the threshold stress \(F_0 = 1\).

The dynamics include the following two steps. (i) Driving: We pick the site with the largest value of stress \(F_{\rm max}\) and subtract this from the threshold to find the constant rate \(v = F_0 - F_{\rm max}\). Then, we increase the stress of all sites simultaneously as 
\begin{equation}
	F_i \to F_i + v. \nonumber
\end{equation}
(ii) Relaxation: The uniform driving causes a site to become unstable \(F_i\geq F_0\). The unstable sites relax in parallel by the following update rules:  
\begin{eqnarray}
	F_i \to 0, \nonumber \\ F_j \to F_j + \alpha F_i, \nonumber
\end{eqnarray}
where \(j\) denotes nearest-neighbors of the site \(i\) on the square lattice. \(\alpha = 1/q = 0.25\) denotes locally conservative dynamics, while \(\alpha<0.25\) corresponds to locally dissipative dynamics on the square lattice with the coordination number \(q = 4\). The dissipation occurs at boundaries and locally if \(\alpha<0.25\). The relaxation events resemble a cascade on a fast time scale, meaning that during this, no further driving occurs. The relaxation events stop when all sites attain a stress value smaller than the threshold.

The relaxation event forms an avalanche with size \(S\), the number of relaxed unstable sites, and area \(A\) represents the spatial extent or the number of distinctly discharged sites. The time to complete a relaxation event is the duration \(T\).
Repeatedly, we uniformly drive the system and apply the same relaxation  rule discussed above.

\begin{figure}[t]
	\centering
	\scalebox{1}{\includegraphics{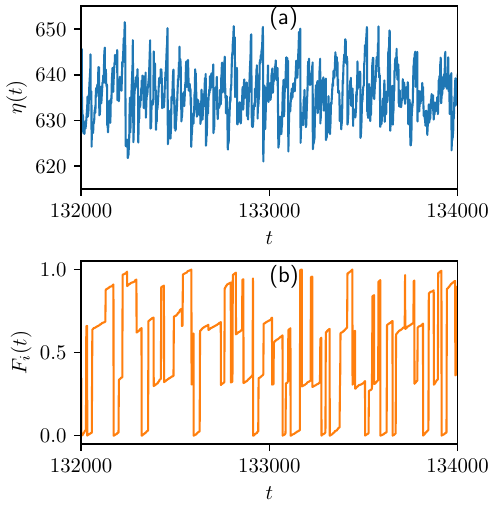}}
	\caption{For the OFC model with \(\alpha=0.25\): (a) A typical signal for the total stress fluctuations \(\eta(t)\) with a system size \(L = 2^6\). (b) A portion of the local stress fluctuations.}
	\label{fig_psd_ft_0}
\end{figure}

\begin{figure}[t]
	\centering
	\scalebox{1}{\includegraphics{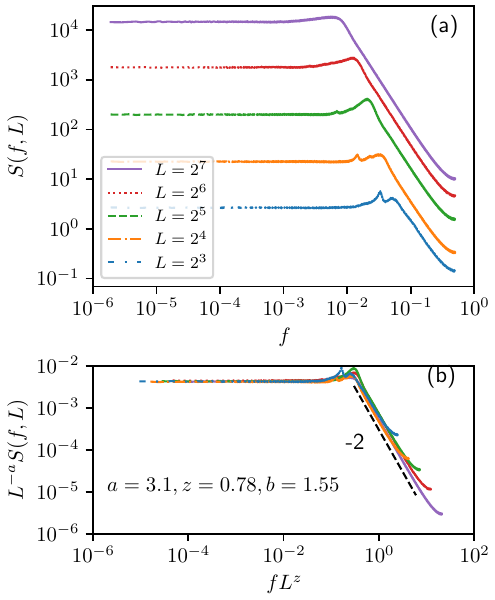}}
	\caption{The OFC model with \(\alpha=0.25\): (a) The power spectra for the total stress fluctuations for different system sizes. The signal length is \(2^{20}\) after discarding transients of \(10^5\). We average each curve over \(10^4\) independent realizations of the process. (b) The data collapse of the power spectra.}
	\label{fig_psd_ft_1}
\end{figure}

\begin{figure}[t]
	\centering
	\scalebox{1}{\includegraphics{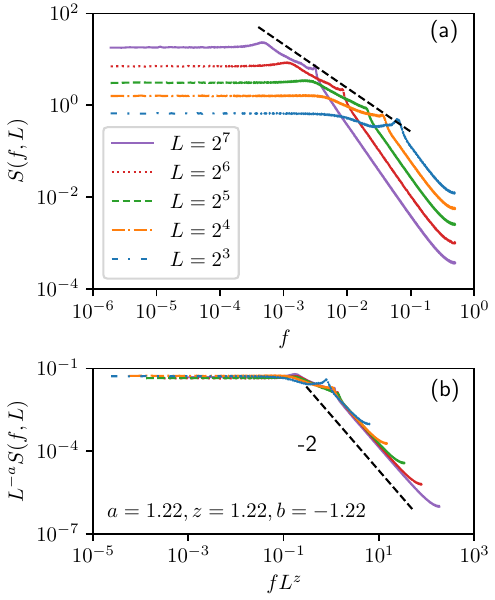}}
	\caption{The OFC model with \(\alpha=0.25\): (a) The power spectra for the local stress fluctuations for different system sizes. The dashed line in the intermediate frequency regime has a slope value of -0.95. (b) The data collapse of the power spectra.}
	\label{fig_psd_fl_2}
\end{figure}

\begin{figure}[t]
	\centering
	\scalebox{1}{\includegraphics{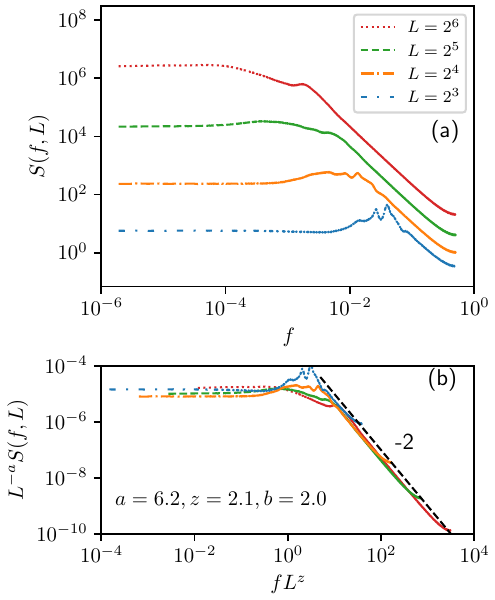}}
	\caption{Same as Fig.~\ref{fig_psd_ft_1} for the total stress noise at \(\alpha = 0.21\).}
	\label{fig_psd_q_ft_1}
\end{figure}

\begin{figure}[t]
	\centering
	\scalebox{1}{\includegraphics{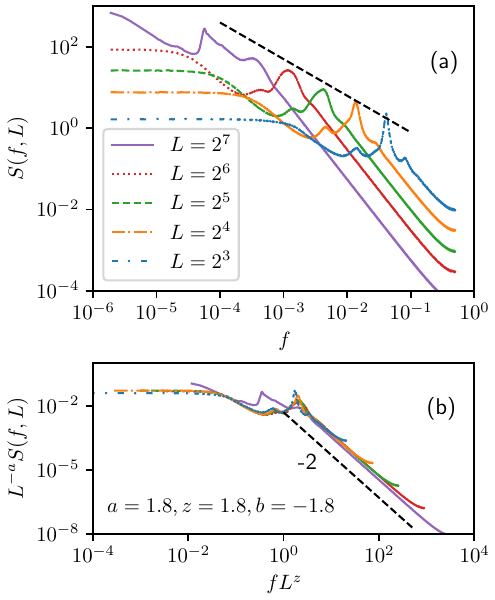}}
	\caption{Same as Fig.~\ref{fig_psd_fl_2} for the local stress noise at \(\alpha = 0.21\). The straight line in (a) has a slope value of 0.9. }
	\label{fig_psd_q_fl_2}
\end{figure}

\section{Power spectra of the stress fluctuations}{\label{sec_3}}
We simulate the OFC model on a square lattice following an efficient algorithm (discussed in Appendix~\ref{app_1}). In this Section, we examine power spectra for the total and local stress or force fluctuations to understand the underlying temporal correlation function in the OFC model for two cases: (i) with locally conservative dynamics \(\alpha = 0.25\) and (ii) with locally dissipative dynamics \(\alpha = 0.21\). Previous studies revealed that \(\alpha = 0.21\) represents a physically relevant case. Figure~\ref{fig_psd_ft_0} shows a typical signal in the statistically steady state for the local stress noise \(F_i(t)\) and the total stress process \(\eta(t) = \sum_iF_i(t)\) with \(i\) running from 1 to \(L^2\).

We can numerically compute the power spectrum of a statistically stationary signal \(x_L(t)\) of length \(M\) as 
\begin{equation}
	\nonumber
	S(f, L) = \lim _{M\to \infty} \frac{1}{M}  \langle |\tilde{x}_L(f)|^2\rangle,
\end{equation}
where \(\tilde{x}_L(f)\) denotes the Fourier transform of the signal, given by  
\begin{equation}
	\tilde{x}_L\left(f = \frac{k}{M}\right) =  \sum_{t=0}^{M-1}x_L(t) \exp\left( -i2\pi \frac{k}{M}t\right),\nonumber
	\label{eq_ps_x1}
\end{equation}
with \(k = 0, 1, 2, \cdots, M-1\). The relevant frequency window is \(f \in [1/M, 1/2]\), with the lowest-frequency bin \(\Delta f = 1/M\). As the power can show an explicit system-size dependence, we prefer not to normalize the power spectrum. We use the standard Fast Fourier Transform algorithm.

For the OFC model with locally conservative dynamics, Fig.~\ref{fig_psd_ft_1}(a) presents power spectra \(S(f, L)\) of the total stress fluctuations \(\eta(t)\) with different system sizes \(L\). On the double logarithmic scale, the power spectrum at a fixed system size exhibits two frequency regimes: (i) Below a lower cutoff frequency, the power remains constant, implying a lack of temporal correlation beyond a correlation time. (ii) Above the lower cutoff frequency, the power spectrum shows a trivial scaling behavior as \(\sim 1/f^2\). However, when the system size increases, we note a nontrivial system size scaling in both frequency regimes as \(S(f\to 0, L) \sim L^a\) and \(S(f\gg f_0, L) \sim L^b\), where the cutoff frequency also scales with the system size \(f_0 \sim L^{-z}\). The scaling features allow us to write 
\begin{equation}
	S(f, L) \sim \begin{cases}L^a, ~~~~~~~~{\rm for}~~~ f\ll L^{-z},\\ L^b/f^{\beta},~~~{\rm for}~~~ L^{-z}, \ll f\ll 1/2. \end{cases}
	\label{eq_psd_1}
\end{equation} 
In terms of the reduced frequency \(u \sim fL^z\), we can also express Eq.~(\ref{eq_psd_1}) as a scaling ansatz 
\begin{equation}
	S(f, L) \sim L^aH(fL^z),
	\label{eq_psd_2}
\end{equation} 
where the scaling function behaves as constant for \(u\ll1\) and \(H(u) \sim u^{-\beta}\) for \(u\gg1\), demanding a scaling relation 
\begin{equation}
	\beta = (a-b)/z.
	\label{eq_psd_3}
\end{equation}  
Equation~(\ref{eq_psd_2}) suggests that we can obtain data collapse for the power spectra by determining two exponents \(a\) and \(z\). By tuning the critical exponents, we find the scaling function as shown in Fig.~\ref{fig_psd_ft_1}(b). The estimated values of the critical exponents \(a = 3.1\), \(b = 1.55\), and \(z = 0.78\) yield the spectral exponent \(\beta \approx 2.0\) with the aid of scaling relation [cf. Eq.~(\ref{eq_psd_3})].

Similarly, we examined the power spectra of the local stress fluctuations for the OFC model with \(\alpha =0.25\) as shown in Fig.~\ref{fig_psd_fl_2}(a). Here, we observe three distinct frequency regimes at a fixed system size. (i) Below a lower cutoff frequency, the power remains constant. (ii) In the high-frequency regime, we find a trivial \(\sim 1/f^2\)-type scaling behavior. (iii) There also exists an intermediate frequency regime where the power spectrum shows a nontrivial spectral exponent close to 1, a signature of canonical \(1/f\) noise. Again, the power spectra show system-size scaling features in all the frequency regimes. While the power grows as \(S(f\to 0,L) \sim L^a\) below the lower cutoff \(f_0 \sim L^{-z}\), the power decays as \(S(f\gg f_0,L) \sim L^{-b}\) in the high frequency regime. For the local stress power spectra, Eqs.~(\ref{eq_psd_2}) and (\ref{eq_psd_3}) remain valid with a negative value of \(b\). Figure~\ref{fig_psd_fl_2}(b) shows the numerically found scaling function for the local stress power spectra.
The estimated values of the critical exponents \(a = 1.22\), \(b = -1.22\), and \(z = 1.22\) yield the spectral exponent \(\beta \approx 2.0\) [cf. Eq.~(\ref{eq_psd_3})].

We also examine the stress power spectra for the locally dissipative OFC model with \(\alpha = 0.21\). As shown in Fig.~\ref{fig_psd_q_ft_1} and \ref{fig_psd_q_fl_2}, we observe qualitatively similar behavior, except in the intermediate frequency regime. However, the FSS yields reasonable data collapse for both local and total stress power spectra. We find that quantitively, the critical exponents \(a\), \(b\), and \(z\) take different values. For the total stress signal, the dynamic exponent increases from 0.8 (with \(\alpha =0.25\)) to 2.1 (with \(\alpha =0.21\)). Similarly, for the local stress signal, the dynamic exponent increases from 1.22 (with \(\alpha =0.25\)) to 1.8 (with \(\alpha =0.21\)). The spectral exponent remains unchanged for both quantities, independent of the dissipation parameter \(\alpha\).

It appears that \(1/f^2\)-type scaling of the local stress contributes dominantly to the total stress fluctuations power spectra. However, the local cross-spectra seem to yield a nontrivial FSS in the power spectrum.

\section{Probability distribution for the difference of avalanche size and area}
\begin{figure}[t]
	\centering
	\scalebox{1}{\includegraphics{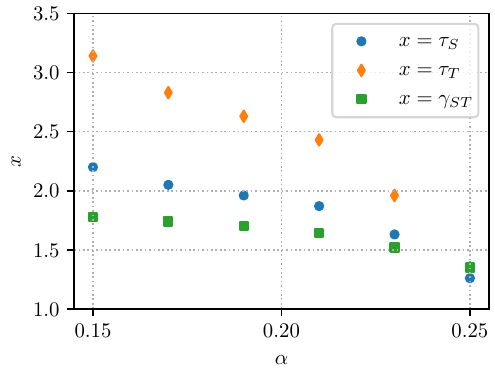}}
	%\scalebox{1}{\includegraphics{fig_av_ofc_exp_2.pdf}}
	\caption{The OFC model on a square lattice: The critical exponents characterizing avalanche properties in the OFC model on a square lattice with \(\alpha\), a parameter capturing the level of local dissipation. When \(\alpha=0.25\), the dynamics are locally conservative. \(\tau_S\) and \(\tau_T\) are the avalanche size and duration exponents, and \(\gamma_{ST} = (\tau_T-1)/(\tau_S-1)\) describes scaling behavior of average size with duration. When the local dissipation increases (or \(\alpha\) decreases), \(\gamma_{ST} \) increases. We also note that \(\tau_A \approx \tau_S\), where \(\tau_A\) is the avalanche area exponent.}
	\label{fig_pd_exp}
\end{figure}

\begin{figure}[t]
	\centering
	\scalebox{1}{\includegraphics{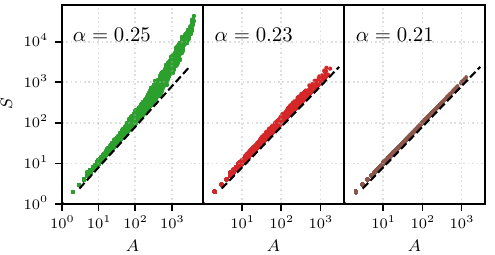}}
	\caption{The avalanche size \(S\) versus area \(A\) for different values of the local dissipation parameter \(\alpha\). The dashed line shows \(S=A\). For \(\alpha=0.25\), one can observe \(S\) is not linearly related to \(A\) and decreasing \(\alpha\) reduces the spread in \((S-A)\).}
	\label{fig_sa}
\end{figure}

\begin{figure}[t]
	\centering
	\scalebox{1}{\includegraphics{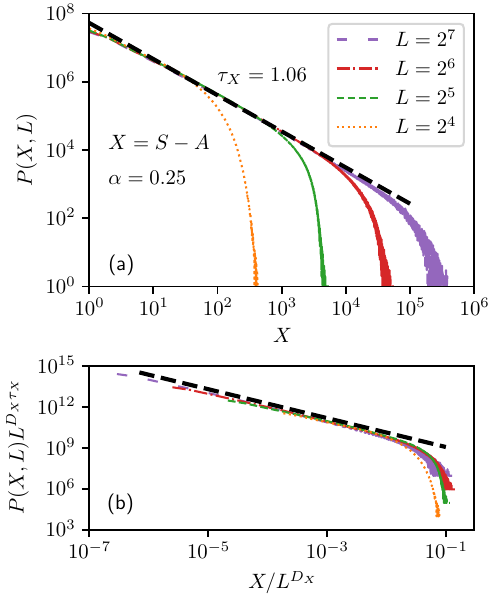}}
	\caption{The OFC model: (a) The probability distribution for the size area difference \(X = S-A\) for different system sizes \(L\) with locally conservative OFC dynamics \(\alpha=0.25\). The total number of avalanches is \(10^9\). About 61.25\% of events for \(L^7\) have \(X = 0\) or the same size and area values. The dashed straight line guides the slope value. (b) The data collapse plot with \(D_X = 3.1\) and \(D_X\tau_X = 3.3\). The estimated value of the size area difference exponent is \(\tau_X = 1.06\).}
	\label{fig_pd_x_1}
\end{figure}

\begin{figure}[t]
	\centering
	\scalebox{1}{\includegraphics{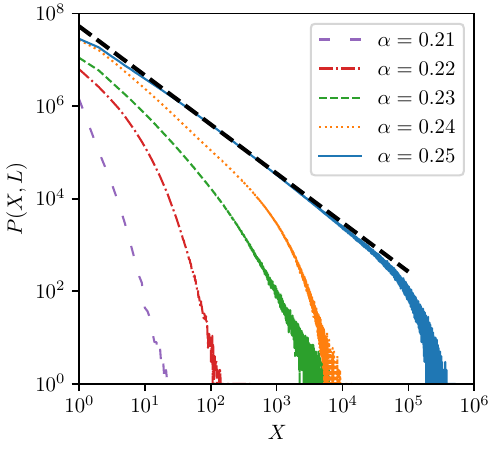}}
	\caption{Same as Fig.~\ref{fig_pd_x_1}(a) but with different values of the dissipation parameter \(\alpha\). The system size is \(L = 2^7\). Notice that the power-law scaling feature of \(X\) disappears when the local dissipation level drastically increases.}
	\label{fig_pd_x_2}
\end{figure}

{\label{sec_4}}
Notice that size and area follow a power-law probability distribution with nearly equal exponents as a function of the dissipation parameter (cf. Fig.~\ref{fig_pd_exp}). Here, we shall examine the difference between avalanche size and area \(X = S-A\) that captures the influence of the affected spatial region because of the avalanche event. This quantity remains unreported, although it is of significant interest in earthquakes. \(X = 0\) when the avalanche size equals the area. The size-area plot shown in Fig.~\ref{fig_sa} reveals the spread that reflects a nonlinear scaling relationship. We find that the difference between size and area reveals interesting scaling features, as discussed below. Figure~\ref{fig_pd_x_1}(a) shows the probability distribution of \(X\) for different system sizes in the OFC model with locally conservative dynamics at \(\alpha = 0.25\). We find a power-law behavior with \(\tau_X = 1.06\) with an upper cutoff. Following Eq.~(\ref{eq_pd_1}), we find a data collapse as shown in Fig.~\ref{fig_pd_x_1}(b). The reasonable data collapse yields \(D_X = 3.1\) and  \(\tau_XD_X = 3.3\). Next, we examine the role of the dissipation parameter \(\alpha\) on \(P(X, L)\). As shown in Fig.~\ref{fig_pd_x_2}, we find that increasing the local dissipation (decreasing \(\alpha\)) drastically results in the loss of the power-law scaling feature for \(X\).

\begin{figure}[t]
	\centering
	\scalebox{1}{\includegraphics{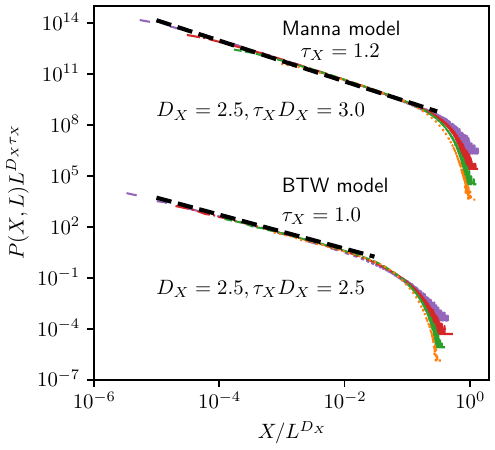}}
	\caption{Same as Fig.~\ref{fig_pd_x_1}(b) but for the BTW and Manna sandpile models on a square lattice. The system size values and the number of avalanches are the same as in Fig.~\ref{fig_pd_x_1}(a).}
	\label{fig_pd_x_3}
\end{figure}

We also examine the extent of the scaling feature in two related SOC systems, namely the BTW and Manna sandpile models on a square lattice [cf. Appendix~\ref{app_2}]. Both models are locally conservative. As shown in Fig.~\ref{fig_pd_x_3}, we find power-law decaying scaling behavior obeying finite-size scaling. The data collapse analysis reveals that \(\tau_X = 1.0\) and \(D_X = 2.5\) for the BTW model and \(\tau_X = 1.2\) and \(D_X = 2.5\) for the Manna model.

\section{summary}{\label{sec_5}}
In summary, we studied the OFC model of earthquakes on a square lattice with open boundary conditions. The model dynamics are locally conservative or non-conservative depending on whether \(\alpha = 0.25\) or \(\alpha < 0.25\), respectively. The avalanche size distribution follows a decaying power law with a non-universal critical exponent, explaining the Gutenberg-Richter law~\cite{gr_1944}. Here, we investigated the local and total stress fluctuations in the OFC model for both locally conservative and non-conservative dynamics. The FSS analysis of the power spectra for the stress fluctuations reveals qualitatively the same but quantitatively significantly different behavior. The dynamic exponent describing the divergence of the correlation time with system size changes from nearly ballistic in the conservative to diffusive dynamics in the locally dissipative dynamics with \(\alpha = 0.21\). The local stress also exhibits a signature of nearly canonical \(1/f\) noise in the intermediate regime, while $1/f^2$-type scaling dominates the high-frequency regime. It would be of interest to examine the extent of such scaling features in the OFC model on a complex network to understand more realistic behavior.

We also examined the probability distribution of the difference between avalanche size and area. This remained previously unexplored, despite being of significant interest in earthquakes~\cite{konstantinou_2014, murotani_2013}. We found a power-law probability with an exponent value of one for the conservative OFC model. The scaling feature vanishes when the local dissipation level increases, even for the physically relevant case \(\alpha = 0.21\). To assess the robustness of such features, we also examined the same quantity in the BTW and Manna sandpile models on a square lattice. We again find that the power-law behavior survives for these systems due to locally conservative dynamics.

The power-law probability distribution behavior of the difference between size and area, $X$, reflects a nonlinear relationship between average size and area scaling. For a physically relevant case, the avalanche size distribution exponent is 1.8 in the OFC model with $\alpha = 0.21$, but we observe that the size-area scaling becomes linear. In this case, the distribution of $X$ does not exhibit a scaling feature. In empirical studies, the size-area scaling exhibits a nonlinear behavior. Our results quantitatively demonstrate the limitations of the OFC model. However, the behaviors are qualitatively consistent with real observations for $\alpha >0.21$. 

\section*{ACKNOWLEDGMENTS}
RC thanks UGC, India, for financial support through the Senior Research Fellowship.

\appendix
\section{Efficient algorithm for the OFC models}{\label{app_1}}
We utilize Grassberger's efficient algorithm to simulate the OFC model~\cite{grassberger_1994}. We convert the \(2D\) array into \(1D\) array by storing the sites information in \(1D\) array. Now, one can access any site by using only 1 index, and it is easy to precompute the nearest neighbors of all sites. During the parallel update, we do not check the entire lattice, but we only keep track of the currently active sites. It significantly improves the performance of the simulation.

\section{BTW and Manna Sandpile models}{\label{app_2}}
BTW introduced the concept of SOC in 1987 to explain \(1/f\) noise~\cite{btw_1987}. They introduced a model, inspired by sandpiles, to show that slow driving and fast dissipation can generate scale-free avalanches in the system. The BTW model has a medium defined as a square lattice with open boundary conditions. Each site has a height variable \(h\) that can take discrete values, representing sand particles. At each time step, one particle is dropped on a randomly selected site \(i\) by increasing \(h_i\) by one unit. We fix a critical threshold \(h_c = 4\) such that if \(h_i \ge h_c\) then the site become active and topples as
\begin{eqnarray}
	h_i \to h_i - 4, \nonumber \\ h_q \to h_q + 1, \nonumber
\end{eqnarray}
where \(q\) runs over the nearest neighbors of the randomly selected site. This step can activate other nearby sites, which will also topple, generating a cascade event. This is known as an avalanche event, and the external drive is paused until \(h<h_c\) for all sites. The total number of sites and distinct sites that topple during an avalanche event are the avalanche size and area, respectively.

In the BTW model, the local update rules are deterministic. Manna introduced a variant of the sandpile model with \(h_c=2\)~\cite{Manna_1991}. The active site topples, and 2 particles are removed from the site. Manna suggested including stochasticity in the redistribution rule by randomly and independently selecting two neighbors and giving one particle to each site~\cite{Manna_1991, sandpile_2025}. Notice that the same neighbor can be selected twice in this process. The BTW model does not obey FSS, while the Manna model does. The BTW and Manna models are different from the OFC model in the following ways~\cite{ghaffari_1997}: (i) the height of the toppled site is reduced by an amount \(h_c\) in sandpile models, but it is set to zero in the OFC model. (ii) Sandpile models follow random driving, but the OFC model has uniform driving, \ie height of all the sites is increased until any one site becomes active.

\bibliography{paper}
\bibliographystyle{myrev}

\end{document}